# A Reactive Molecular Dynamics Model for Uranium/Hydrogen Containing Systems.

*Artem Soshnikov,[1] Rebecca Lindsey,[2] Ambarish Kulkarni,[1] Nir Goldman[1,3]\**

[1]Department of Chemical Engineering, University of California, Davis, CA 95616, USA

[2]Department of Chemical Engineering, University of Michigan, Ann Arbor, MI 48109, USA

[3]Physical and Life Sciences Directorate, Lawrence Livermore National Laboratory, Livermore, CA, 94550 USA.

\*Email: goldman14@llnl.gov

Topics: hydriding, uranium, machine learning, Density Functional Theory, corrosion, embrittlement, computational models

**ABSTRACT**

Uranium-based materials are valuable assets in the energy, medical, and military industries. However, understanding their sensitivity to hydrogen embrittlement is particularly challenging due to the toxicity of uranium and computationally expensive nature of the quantum-based methods generally required to study such processes. In this regard, we have developed a Chebyshev Interaction Model for Efficient Simulation (ChIMES) model that can be employed to compute energies and forces of U and $UH_3$ bulk structures with vacancies and hydrogen interstitials with similar accuracy to Density Functional Theory (DFT) while yielding linear scaling and orders of magnitude improvement in computational efficiency. We show that that the bulk structural parameters, uranium and hydrogen vacancy formation energies, and diffusion barriers predicted by the ChIMES potential are in strong agreement with the reference DFT data. We then use ChIMES to conduct molecular dynamics simulations of the temperature-dependent diffusion of a hydrogen interstitial and determine the corresponding diffusion activation energy. Our model has particular significance in studies of actinides and other high-Z materials, where there is a strong need for computationally efficient methods to bridge length and time scales between experiments and quantum theory.

**INTRODUCTION**

Uranium (U) is a unique material with a number of practical uses, such as nuclear fuel for electricity generation, radio-isotope sources for diagnosis and research in the medical industry, and as a power source for submarines and weaponry by the military.[2] The pure metal occurs in three solid polymorphs: α (orthorhombic), β (tetragonal) and γ (body-centered cubic). The most prominent metal phase in nature is α-U, shown in Figure 1a, which transforms to β-U at approximately 935 K and subsequently transforms to γ-U at approximately 1045 K.[3] However, uranium is also a highly reactive metal and under ambient conditions will react spontaneously with hydrogen gas to form a brittle hydride ($UH_3$), causing disintegration of the underlying metal. $UH_3$ itself is pyrophoric, leading to operational hazards and making surface hydrogenation experiments highly challenging.[4] Uranium hydride can exist in two different phases: α-$UH_3$ and β-$UH_3$. Both phases possess a cubic lattice with each U atom is surrounded by 12 H atoms, as shown in Figure 1(b and c). The α-$UH_3$ phase is metastable at low temperature exhibits a face-centered cubic (fcc) symmetry, with one $UH_3$ formula unit per primitive cell or four formula units in the cubic unit cell. In contrast, the ground-state β-$UH_3$ phase exhibits lower symmetry with eight formula units per unit cell, though with cubic symmetry overall. The more compact α-phase completely converts to the β-phase at approximately 375 K, generally below the operating conditions of many nuclear reactors.[5] Relatively little is known about the hydriding mechanism in pure uranium, in part due to the material's reactive nature. Future studies would thus greatly benefit from atomic-level simulations of the hydriding process, which can provide microscopic details about the hydrogen-uranium reaction and help guide future experimentation.

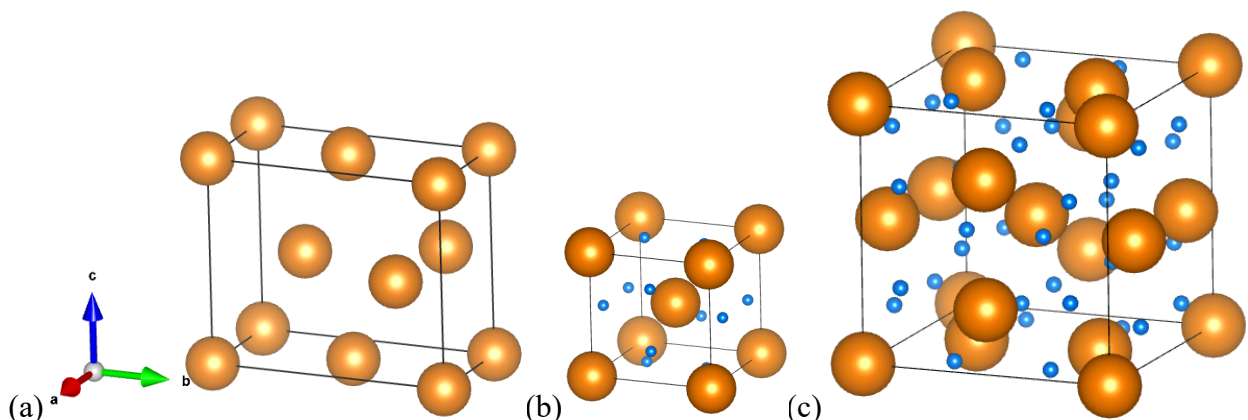

**Figure 1**. Crystal structure of (a) α-U (b) α-UH$_3$, and (c) β-UH$_3$ drawn using Vesta (version 3.0).[1]

In order to probe the formation of new material phases and the formation of covalent bonds, atomistic simulations frequently require use of Kohn-Sham Density Functional Theory (DFT), which has shown immense predictive capability for material phases over a wide range of thermodynamic conditions.[6–9] However, DFT calculations require significant computational resources per simulation step, and therefore are generally limited to time scales on the order of picoseconds and system sizes of few hundreds of atoms. Small-scale U+H calculations are relatively tractable with DFT and can be used in dataset preparation and validation. However, DFT calculations are too computationally intensive to model polycrystalline regions, grain boundaries, and realistic defect concentrations that likely play a significant role in the hydriding process. In fact, for this study, approximately 70,000 CPU-hours was required to run a molecular dynamics (MD) trajectory of a small α-UH$_3$ system (54U + 162H) for only 0.5 ps. In contrast, convergence of hydride initiation, nucleation, and growth studies could require simulation cells of tens of thousands of atoms or greater run for nanosecond timescales or longer.[10] Therefore, uranium hydriding atomistic simulations require an alternative fast, accurate, and computationally-inexpensive approach that can calculate large-scale effects that are computationally cost prohibitive to determine using DFT alone.

A practical solution to ameliorate these system size and time scale limitations is the development of computationally efficient MD force fields that can approximate the underlying potential energy surface with accuracy comparable to DFT. In this respect, classical force field approaches[11–13] have traditionally shown outstanding computational efficiency in modeling materials. These empirical approaches, though, generally do not allow for reactive conditions where bond breaking and forming occurs. Development of reactive force field methods, such as ReaxFF[14] and COMB,[15] incorporate both reactive and non-reactive terms with physically motivated bond-order forms, and allow for bond breaking and forming under realistic conditions. However, these methods frequently involve rigid functional forms that can require potentially complex optimizations of non-linear parameters. More recently, machine learning (ML) approaches for MD simulations have been developed that utilize many-body kernels in more abstract, highly flexible functional forms. Examples include Gaussian Approximation Potential (GAP),[16] which leverages Gaussian Process Regression, and DeepMD,[17,18] which leverages deep neural networks. These ML approaches have shown high degree of accuracy and transferability for a number of systems.[19,20] ML approaches generally require large training and validation datasets for training purposes as well significant training times due to their inherent non-linear optimization requirements. These issues can be of particular challenge for actinide containing systems, where existing DFT data can be limited, and training data can be difficult to generate due to the extreme computational effort associated with quantum calculations of high-Z materials.

Machine-learned methods that rely on linear parameterization, such as the Chebyshev Interaction Model for Efficient Simulation[21,22] (ChIMES), hold promise as an potentially easier to optimize model for accelerated MD simulations with a high degree of accuracy. ChIMES is a many-body reactive force field for MD simulation based on linear combinations of many-body Chebyshev polynomials. The use of linear parameterization allows for optimal fitting coefficients to be solved for directly in most cases as well as powerful regularization approaches which are not necessarily available to non-linear optimization problems. ChIMES is based on an N-body expansion of DFT energies and forces and thus allows for a physically motivated and highly flexible functional form. In addition, use of Chebyshev polynomials imparts several advantages, including: (1) Chebyshev polynomials of the first kind are orthogonal and can be generated recursively, forming a complete basis set, (2) the derivatives of Chebyshev polynomials of the first kind are related to Chebyshev polynomials of the second kind, which are also orthogonal and generated recursively, (3) higher-order polynomials tend to have decreasing expansion coefficient values due to their monic form, and (4) Chebyshev polynomials of the first kind are nearly optimal, which means that the error due to interpolation closely resembles a minimax polynomial. ChIMES model optimization can be determined relatively quickly (e.g. within minutes for each optimization step of our study). In addition, ChIMES models have been shown in some cases to have significantly smaller data requirements and numbers of parameters than some neural network approaches,[23] making them ideal for the application space to be studied here. Numerous ChIMES models have been designed for complex systems, such as molten liquid carbon,[21] water,[24,25] high-pressure C/O systems,[26,27] hydrazoic acid ($NH_3$),[28] titanium hydride ($TiH_2$),[29] and silicon.[30]

In this work, we detail our efforts to create a ChIMES model for use in uranium hydriding studies. We start with a brief discussion of our DFT calculations as well as the ChIMES methodology. We then investigate different options for optimal values the ChIMES hyperparameters, including polynomials orders for different bodied interactions, the minimum and maximum interatomic distance cutoffs, and regularization parameters. We validate our model against computational and experimental results, including lattice constants and the bulk moduli of different U-H phases, as well as defect energies for single and multiple defects of uranium vacancies, hydrogen interstitials, and hydrogen vacancies in uranium hydride. Finally, we present results from simulations with our optimum model, including the kinetic properties for bulk hydrogen diffusion through bulk α-U, and molecular dynamics simulation of diffusion coefficients as a function of temperature. In all cases, we find that ChIMES yields a high degree of accuracy relative to DFT calculations on smaller system sizes.

## COMPUTATIONAL METHODS

### a) DFT

All of our DFT calculations were performed using the Vienna ab initio simulation package code (VASP)[31–33] with the projector augmented wave (PAW) pseudopotentials[34,35] for U and H with the Perdew–Burke–Ernzerhof generalized gradient approximation (PBE) exchange-correlation functional.[36] In terms of model optimization, we choose to focus on the high symmetry α-UH$_3$ phase, and leave results from β-UH$_3$ for validation. The energy cutoff for the planewave basis set was set to 500 eV based on convergence tests. Structural relaxations were performed until forces on all atoms were less than 0.01 eV/Å. A k-point mesh of 4x4x4 generated by the Monkhorst-Pack[37] method for integration over the Brillouin zone was used to generate the training set for both α-U and α-UH$_3$, discussed below.

Our full set of reference data contains 792 snapshots from the following DFT calculations: (1) short (0.5-1 ps) molecular dynamics simulations of α-U and α-UH$_3$ metallic systems at temperatures of 400 K and 1000 K and various hydrogen concentrations, (2) single-point calculations of isotropically distorted α-U and α-UH$_3$ lattices, and (3) supercell structures with uranium vacancies and images from optimized minimum energy path calculations of hydrogen diffusion in α-U. Pure uranium optimization calculations (atomic coordinates and lattice parameters) and DFT-MD simulations of interstitial-containing systems were performed on 4x2x3 supercell with initial lengths 11.209 x 11.687 x 14.711 Å. The interstitial simulations contained 94-96 uranium atoms and various hydrogen concentrations of hydrogen atoms (1-10 H atoms). We used a cubic supercell for defect-free α-UH$_3$, with a lattice vector length of 12.363 Å and containing 54 U and 162 H atoms. All DFT-MD simulations were performed in the canonical ensemble (NVT) with timestep of 4.0 fs for pure systems of uranium and 0.20 fs for systems containing H, using Nose-Hoover thermostat chains[38–40] and periodic boundary conditions. For our training, uniformly spaced frames from the MD calculations were extracted every 150-200 fs in order to ensure that configurations were as statistically uncorrelated as possible.

### b) ChIMES

A detailed explanation of the ChIMES interaction model has been discussed elsewhere[29,30,41] and is summarized here, briefly. The design philosophy behind ChIMES comprises of mapping the DFT total energy onto linear combinations of many-body Chebyshev polynomials of the first kind. The ChIMES total energy is expressed as follows:

$$E_{\text{ChIMES}} = \sum_{i_1}^{n_a} {}^1E_{i_1} + \sum_{i_1 > i_2}^{n_a} {}^2E_{i_1 i_2} + \sum_{i_1 > i_2 > i_3}^{n_a} {}^3E_{i_1 i_2 i_3} + \sum_{i_1 > i_2 > i_3 > i_4}^{n_a} {}^4E_{i_1 i_2 i_3 i_4}$$
$$+ \text{ higher bodied terms.}$$

(1)

Here, ${}^1E_{i_1}$ corresponds to the one-body energies or atomic energy constants, ${}^2E_{i_1 i_2}$ to the two-body energies or pair-wise interactions, ${}^3E_{i_1 i_2 i_3}$ to the three-body energies or triplet interactions, and ${}^4E_{i_1 i_2 i_3 i_4}$ to the four-body energies or quadruplet interactions. The index $n_a$ corresponds to the number of atoms in the system. In ChIMES, each of the greater than one-body terms is expressed as a polynomial sum. For example, the two-body term $^2E_{i_1 i_2}$ is expressed as:

$$^2E_{i_1 i_2} = f_p(r_{i_1 i_2}) + f_c^{e_{i_1} e_{i_2}}(r_{i_1 i_2}) \sum_{m=1}^{O_2} C_m^{e_{i_1} e_{i_2}} T_m\left(s_{i_1 i_2}^{e_{i_1} e_{i_2}}\right)$$

(2)

The indices $\{e_{i_1}, e_{i_2}\}$ correspond to the element types of atoms $i_1$ and $i_2$ respectively. In this case, all pairwise distances $r_{i_1 i_2}$ are transformed over $[r_{min}^{e_{i_1} e_{i_2}}, r_{max}^{e_{i_1} e_{i_2}}]$ (e.g., the range of minimum and maximum values for a given element pair) to the scaled coordinate $s_{i_1 i_2}^{e_{i_1} e_{i_2}}$, which is restricted to the Chebyshev polynomial input variable range of [-1,1]. The function $f_c^{e_{i_1} e_{i_2}}(r_{i_1 i_2})$ assures smooth variation of the energy function at the maximum distance boundary. The function $f_p(r_{i_1 i_2})$ is a penalty function that adds extra repulsion for situations where $r_{i_1 i_2} < r_{min}^{e_{i_1} e_{i_2}}$, (i.e., when the pairwise distance falls below the range of allowable inputs for the Chebyshev polynomials). The two-body sum is performed over all $m$ dimers that exist within the $r_{max}^{e_{i_1} e_{i_2}}$ cutoff, and $\{C_m^{e_{i_1} e_{i_2}}\}$ are the set of optimized fitting coefficients that are permutationally invariant for each pair of atom types.

Higher bodied orthogonal polynomials for clusters greater than a dimer can be constructed by taking the tensorial product of the sum of the constituent $\binom{n}{2}$ unique pairwise polynomials of a that cluster. For example, a triplet or three-body cluster with the set of indices of $\{i_1, i_2, i_3\}$ contains $\binom{3}{2} = 3$ unique distances, namely $r_{i_1 i_2}, r_{i_1 i_3}, r_{i_2 i_3}$. Thus, a three-body polynomial of total order $m + n + q$ is constructed by first applying the transforms $r_{i_1 i_2} \rightarrow s_{i_1 i_2}^{e_{i_1} e_{i_2}}$, $r_{i_1 i_3} \rightarrow s_{i_1 i_3}^{e_{i_1} e_{i_3}}$, $r_{i_2 i_3} \rightarrow s_{i_2 i_3}^{e_{i_2} e_{i_3}}$ and then taking the product $T_m\left(s_{i_1 i_2}^{e_{i_1} e_{i_2}}\right) T_p\left(s_{i_1 i_3}^{e_{i_1} e_{i_3}}\right) T_q\left(s_{i_2 i_3}^{e_{i_2} e_{i_3}}\right)$. The three-body energy $E_{ijk}$ can then be computed as the following linear combination:

$$^3E_{i_1 i_2 i_3} = f_c^{e_{i_1} e_{i_2}}(r_{i_1 i_2}) f_c^{e_{i_1} e_{i_3}}(r_{i_1 i_3}) f_c^{e_{i_2} e_{i_3}}(r_{i_2 i_3}) \sum_{m=0}^{O_3} \sum_{p=0}^{O_3} \sum_{q=0}^{O_3*} C_{mpq}^{e_{i_1} e_{i_2} e_{i_3}} T_m\left(s_{i_1 i_2}^{e_{i_1} e_{i_2}}\right) T_p\left(s_{i_1 i_3}^{e_{i_1} e_{i_3}}\right) T_q\left(s_{i_2 i_3}^{e_{i_2} e_{i_3}}\right) .$$

(3)

In this case, the set of $\{C_{mpq}^{e_{i_1} e_{i_2} e_{i_3}}\}$ correspond to the three-body fitting coefficients that are permutationally invariant to atom types in the set $\{e_{i_1}, e_{i_2}, e_{i_3}\}$ as well as polynomial order. We also apply smoothly varying cutoff functions to the three-body interactions, though penalty functions are omitted in this case and only included in the two-body energies. Finally, the '*' in the sum in Equation (3) corresponds to the fact that we only include distinct triplets in the sum where $r_{i_1 i_2}, r_{i_1 i_3}$, and $r_{i_2 i_3}$ are all less than $r_{max}^{e_{i_1} e_{i_2}}$.

Greater than three-body terms in the ChIMES energy expression are included in an equivalent manner. In practice a maximum of four-body terms are used due to prevent creating a combinatorically large polynomial space and potential parameter explosion.[23,29,30] ChIMES bears some resemblance to other polynomial expansion methods such as the Atomic Cluster Expansion[42,43] (ACE) and spectral neighbor analysis potential[44] (SNAP) methods. We note that the polynomial basis sets in these methods are atom-centered and are functionally different than the cluster-centered Chebyshev approach we employ here.

Similar to other atomic interaction potentials,[45–48] ChIMES models are trained through matching forces, energies, and stress tensor components. In general, training and validation data are generated through DFT optimized structures and MD simulations, though the possibility exists to include data from higher levels of theory. In addition, use of weights is frequently required due to the differing physical units of the forces, stresses, and energies and the number of data points per configuration. We can thus define an objective function for our optimization as follows:

$$F_{obj} = \frac{1}{N_d} \sum_{\tau=1}^{M} \left( \sum_{i=1}^{N_\tau} \sum_{\alpha=1}^{3} \left(w_F \Delta F_{\tau \alpha_i}\right)^2 + \sum_{\alpha=1}^{3} \sum_{\beta \leq \alpha} \left(w_\sigma \Delta \sigma_{\tau \alpha \beta}\right)^2 + (w_E \Delta E_\tau)^2 \right).$$

(4)

The index $\tau$ corresponds to a specific training set configuration from the total set of $N_\tau$ configurations, $i$ is the atomic index, and $\alpha$ and $\beta$ are the cartesian directions. We use the index $M$ to denote the total number of configurations in the training set, with $N_d$ corresponding to the total number of data points (e.g., forces, stress tensor components, and energies). In addition, $\Delta F_{\tau \alpha_i} = F_{\tau \alpha_i}^{ChIMES} - F_{\tau \alpha_i}^{DFT}$, $\Delta \sigma_{\tau \alpha \beta} = \sigma_{\tau \alpha \beta}^{ChIMES} - \sigma_{\tau \alpha \beta}^{DFT}$, $\Delta E_\tau = E_\tau^{ChIMES} - E_\tau^{DFT}$. The value $w_F$ is the weight for forces, $w_\sigma$ for the stress tensor components, and $w_E$ for the energies.

c) **Optimization and Regularization**

Regularization is an important concept that is utilized in order to avoid overfitting of the trained data, and we refer to previous publications for further details.[30] In this work, we use the Least Absolute Shrinkage and Selection Operator method (LASSO), an L1 regularization technique which adds a penalty proportional to the absolute value of the magnitude of the fitting coefficients. This promotes smaller valued coefficients to become zero and subsequently be eliminated from the model. In this case, the objective function $F_{obj}$ is minimized with the following additional penalty on the sum of the absolute values of the fitting coefficients $C_i$:

$$F'_{obj} = N_d F_{obj} + 2\alpha' \sum_{i=1}^{N_p} |C_i|$$

(5)

where $\alpha'$ is the parameter that regularizes the magnitude of the fitting coefficients $C_i$, and $N_p$ is the total number of unique fitting parameters. In our work, we use LASSO as implemented within the Least-Angle Regression (LARS) optimization method,[49] which is discussed in more detail in Refs.[29,30]

## RESULTS AND DISCUSSION

### A. Finding optimal parameters

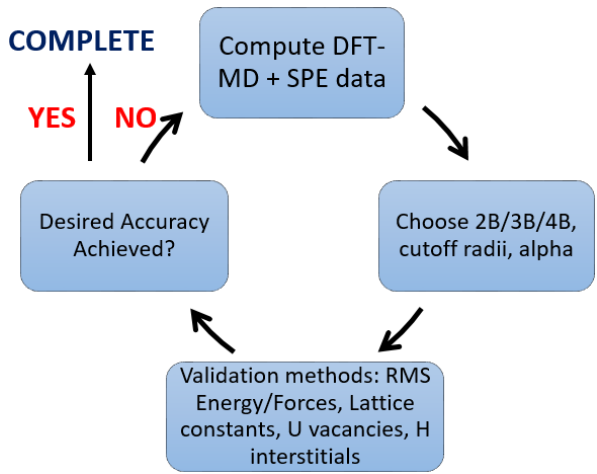

**Figure 2**. Flowchart for creation of ChIMES model.

ChIMES model development requires the definition of a number of *hyperparameters*, i.e., user-chosen model parameters. These include the two, three, and four-body polynomial orders, minimum and maximum atomic interaction distance cutoffs ($r_{min}$ and $r_{max}$), and the regularization method and degree of regularization. Figure 2 shows a workflow diagram for ChIMES model optimization, which comprises exploring different combinations of the fitting parameters. Validation was determined through calculation of the root-mean-square (RMS) error of different physical properties that were not included in our fits, such as the bulk lattice constants, ground-state volumes for α-U and α-UH$_3$ systems, single uranium vacancy formation energy in pure U, and single hydrogen absorption interstitials in an α-U supercell.

**1.1 Optimal interatomic interaction distances.** The $r_{min}$ value for each atomic pair (H-H, U-H, and U-U) was determined by scanning the data and finding the minimum interatomic interaction distances in our dataset (0.50 Å for H-H, 0.68 Å for U-H, and 1.07 Å for U-U pairs). The $r_{max}$ atomic pair values were uniformly set to the maximum allowed distance sampled in our training set (5.5 Å) in order to satisfy the minimum image convention (i.e., one half of the shortest box length).

**1.2 Sweep of polynomial order.** Given our choice of $r_{min}$ and $r_{max}$, we have created multiple models with our training set by sampling a range of polynomial orders for the two-body (2B), three-body (3B), and four-body (4B) interactions. Here, we have looped over our workflow by sampling the 2B order a range of $2 \leq O_{2B} \leq 18$, the 3B order over a range of $2 \leq O_{3B} \leq 14$, each with a step size of two. We have also created a subset of potentials with the 4B order varied over range of $1 \leq O_{4B} \leq 4$, in steps of one. Our preliminary studies showed that models with 2B only interactions produced substantially diminished results for all validation tests. Therefore, in this work we discuss results from models with combinations of non-zero 2B/3B order values, only. The optimizations were performed with the LASSO regularization of $\alpha = 10^{-3}$, similar to previous work.[29,30] In doing so, we are able to validate a large number of independent models and eventually down select to our optimal choice. In this section, we first present results for models with 2B/3B polynomial models, followed by studies with incorporation of the 4B terms.

The results for the computed the root-mean-square (RMS) error in our training set for the atomic forces on hydrogen and uranium ions are shown in Figure 3. A more detailed RMS error results summary, which includes diagonal of the steps tensor and energies, are shown in the Supplementary Materials section. Unsurprisingly, the RMS error for forces of each fit decreases systematically with higher polynomial orders and higher bodied models. For example, a model with a set of $\{O_{2B}=8, O_{3B}=6\}$ yields fitting errors of 0.489 and 1.971 eV/Å for the forces acting on H and U, respectively. On the other hand, a model with a set of $\{O_{2B}=18, O_{3B}=12\}$ produces substantially lower fitting errors of 0.197 and 0.342 eV/Å. We notice similar trends for RMS errors of the diagonal of the stress tensor (e.g. 1.962 GPa for $\{O_{2B}=8, O_{3B}=6\}$ compared to 0.696 GPa for $\{O_{2B}=18, O_{3B}=12\}$ model). On the other hand, the opposite trends are observed for the total energy as increasing polynomial order yields larger RMS values (e.g. 0.283 eV/U atom for $\{O_{2B}=8, O_{3B}=6\}$ and 0.421 eV/U atom for $\{O_{2B}=18, O_{3B}=12\}$ sets). We also note that models with higher orders of polynomials require more computational resources for the optimization process (e.g., about ten times the computational effort for the $\{O_{2B}=18, O_{3B}=12\}$ model discussed here), though the optimization remains relatively rapid (approximately 45 minutes on a single Intel Xeon E5-2695 processor for the above example).

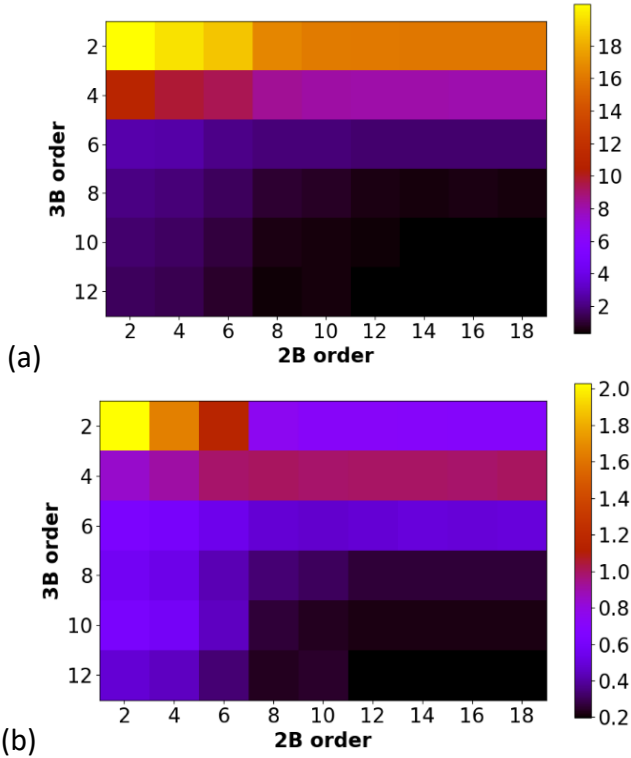

**Figure 3**. Results for training RMS errors using $O_{2B}$ and $O_{3B}$ Chebyshev polynomials. RMS error in the forces in units of eV/Å for (a) on U atoms and (b) on H atoms.

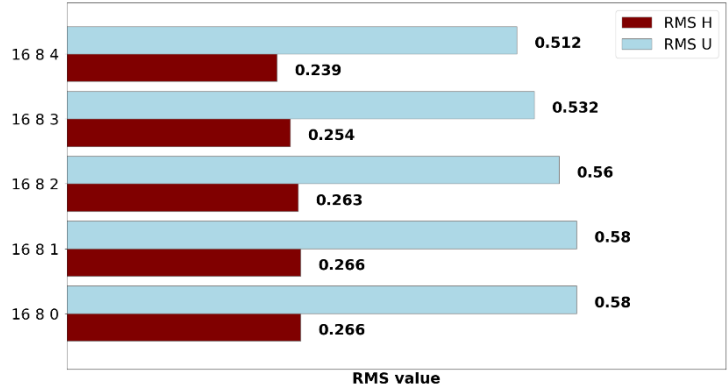

**Figure 4**. Results for training RMS error test with the inclusion of $O_{4B}$ Chebyshev polynomials.

Incorporation of 4B interactions yields further training accuracy, where we have included terms up to $O_{4B}=4$, similar to previous work.[23,29] As shown in Figure 4 and Table S1 (Supplementary Materials), we observe a monotonic decrease in RMS training error with inclusion of higher four-body order in our model. We see RMS force errors of 0.580 eV/Å in U and 0.266 eV/ Å in H for $\{O_{2B}=16, O_{3B}=8, O_{4B}=0\}$, 0.580 eV/Å and 0.266 eV/Å for $\{O_{2B}=16, O_{3B}=8, O_{4B}=1\}$, and 0.560 eV/ Å and 0.263 eV/ Å for $\{O_{2B}=16, O_{3B}=8, O_{4B}=2\}$. In this case, the RMS errors for the total energy diminish with increasing 4B polynomial order (e.g. 0.243 eV/U atom for $O_{4B}=0$, and 0.238 eV/U atom for $O_{4B}=3$ models), though these improvements are small. We notice that ChIMES model with $O_{4B}=0$ yields virtually identical results to those from $O_{4B}=1$, which occurs due to LASSO regularization setting these four-body parameters to zero. We observe some measurable reduction in the RMS errors for models with $O_{4B}=3$ or higher. We calculate the RMS force errors of 0.532 eV/ Å in U and 0.254 eV/ Å in H and the total energy error of 0.216 eV/atom for $\{O_{2B}=16, O_{3B}=8, O_{4B}=3\}$ and 0.512 eV/ Å, 0.239 eV/ Å, and 0.191 eV/U atom for $\{O_{2B}=16, O_{3B}=8, O_{4B}=4\}$. The opposite trends are observed in RMS errors of the diagonal of the stress tensor with increasing four-body order (e.g., 0.771, 0.771, 0.778, 0.789, 0.877 GPa).

Next, we perform a model down select through validation against a series of DFT-computed physical properties. These include the lattice constants, volume of the optimized unit cell of α-U and α-UH$_3$, vacancy formation energy in α-U, and hydrogen interstitial energy formation in α-U. We note that β-UH$_3$ was not part of this initial validation test. As shown in Table S1, the complexity of the model in general reduces the absolute errors relative to DFT reference values. However, some validation tests are more sensitive than others. In addition, the lattice constants and equilibrium volume tests converge to sufficient accuracy without utilizing the four-body interactions. For example, a model with a set of $\{O_{2B}=6, O_{3B}=6\}$ yields absolute errors for lattice constants for α-U and α-UH$_3$ below 5.0%. Increasing the polynomial order above these values leads to the reduction of relative errors in lattice, e.g., a model set of $\{O_{2B}=18, O_{3B}=8\}$ produces errors of 1.4% and 4.15% for α-U and 1.83% and 5.4% for α-UH$_3$ lattice and volume parameters.

A similar trend to a certain extent is observed for the formation energy of the uranium vacancy in 4x2x3 α-U supercell. Here, we define the uranium vacancy formation energy is defined as

$$E_v = E_{(n-1)U} - \left[\frac{n-1}{n}\right]E_{nU},$$

(6)

where $E_{(n-1)U}$, $E_{nU}$, and n are the supercell energies for the defective and perfect systems, and n is the number of uranium atoms, respectively. Models with polynomials orders above 10 for 2B and 8 for the 3B interactions estimate vacancy energy with relative error of ±0.4 eV (or 25% errors) or smaller, relative to the DFT computed value of 1.78 eV, with the best model $\{O_{2B}=12, O_{3B}=8\}$ of 1.80 eV with relative error of 0.9% .

We find that accurate prediction for the hydrogen interstitial in α-U to be one of the most challenging properties in our initial validation set. The interstitial formation energy is defined as

$$E_f = E_{U+nH} - E_U - 1/2\, nE_{H_2},$$

(7)

where $E_{U+nH}$, $E_U$, and n are the supercell energies for the defective and perfect systems, and *n* is the number of hydrogen atoms in the defective supercell, respectively. Here, we have calculated the hydrogen interstitial energy for the most stable site (square-pyramidal). Similar to other validation tests, increasing the polynomial order of 2B and 3B interactions improve the results. For example, a model with a set of $\{O_{2B}=6, O_{3B}=6\}$ yields absolute errors of over 600%, while a model $\{O_{2B}=12, O_{3B}=8\}$ estimates the interstitial energy of 0.22 eV (17% relative error). fairly large polynomial orders (above 12 for 2B and 8 for 3B) are required to achieve relative errors below 10% relative to the DFT computed value of 0.27 eV, with the best model $\{O_{2B}=18, O_{3B}=8\}$ of 0.26 eV with relative error of 1.5% .

Lastly, we looked at the effect of including higher-body interactions on our validation tests. As shown in Table S1, incorporation of the four-body polynomial to the model set significantly improves the lattice constants for both of α-U and α-UH$_3$. For example, adding the 4B terms to the model with a set of $\{O_{2B}=16, O_{3B}=8\}$ somewhat improves relative errors, where we observe values of 2.32% (α-U) and 2.13% (α-UH$_3$) for $\{O_{4B}=0\}$ and $\{O_{4B}=1\}$, 2.29% (α-U) and 1.49% (α-UH$_3$) for $\{O_{4B}=2\}$, 2.23% (α-U) and 1.15% (α-UH$_3$) for $\{O_{4B}=3\}$, and 1.8% (α-U) and 1.43% (α-UH$_3$) for $\{O_{4B}=4\}$. However, the additional 4B complexity yields higher errors for other validation tests. Errors in uranium vacancy formation energies systematically increase from ~0.2 eV (0.9%) to ~0.4 eV (20.8%). Similar trends are observed for the hydrogen interstitial with errors of ~0.05 eV (16.8%) for $\{O_{4B}=0\}$ and $\{O_{4B}=1\}$, ~0.07 eV (24.3%) for $\{O_{4B}=2\}$, ~0.18 eV (67.2%) for $\{O_{4B}=3\}$, and ~5.7 eV (>2000%) for $\{O_{4B}=4\}$. Overall, we find comparable accuracy between the $\{O_{2B}=16, O_{3B}=8, O_{4B}=0\}$ model those with $\{O_{2B}=16, O_{3B}=8, O_{4B} \leq 2\}$ for the solid phase lattice constants and point defect energies in this validation suite. We compute some loss of accuracy for ChIMES models with values of $O_{4B} = 3$ or 4 for defect formation energies. As a result, we choose to proceed with ChIMES models with 2B and 3B interactions only, though the inclusion of non-zero 4B interactions has proven essential for simulations of reactive materials over a broad range of thermodynamic conditions.[28]

We now down select to a ChIMES model $\{O_{2B} = 16, O_{3B} = 8, O_{4B} = 0\}$. We find that this model achieves the correct balance of minimizing the RMS errors our training set while yielding accurate bulk parameters for α-U and α-UH$_3$. In particular, we find that this model is able to achieve accurate results for point defect properties, including hydrogen interstitial and uranium vacancy formation. Hence, we choose to proceed with this model for the remainder of our study.

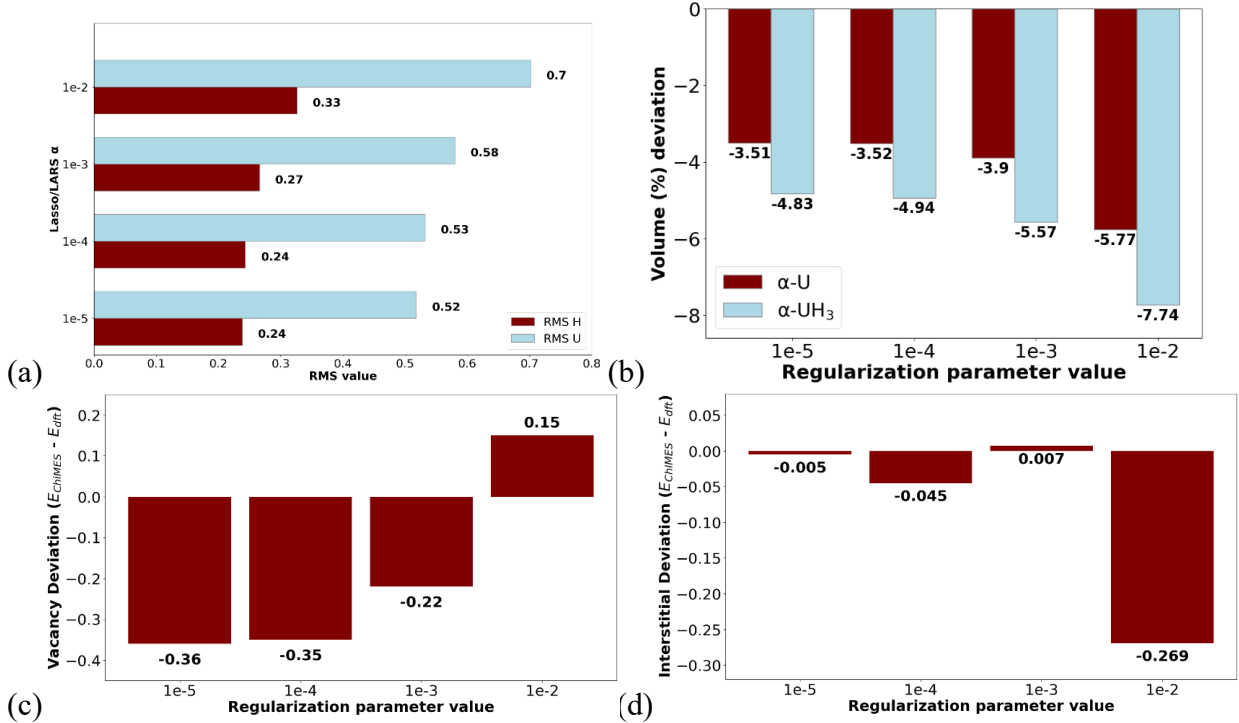

**Figure 5**. Results of validation tests using different values of LASSO/LARS parameter α. (a) Root mean square force errors (in eV/Å) on U and H atoms, (b) Percent deviation in volume of ChIMES estimated optimized unit cell from the DFT value in α-U and α-UH$_3$, (c) Defect energy formation deviation (in eV) of U vacancy in α-U 4x2x3 supercell, and (d) Defect energy formation deviation (in eV) of 1 H interstitial (square-pyramidal) in α-U 4x2x3 supercell.

**1.3 Test of Regularization methods.** Given our choice of the $r_{min}^{H-H} = 0.50$, $r_{min}^{U-H} = 0.68$, $r_{min}^{U-U} = 1.07$ and $r_{max}^{H-H} = 5.5$, $r_{max}^{U-H} = 5.5$, and $r_{max}^{U-U} = 5.5$ atomic interaction distance cutoffs and the ChIMES polynomial order {$O_{2B}$=16, $O_{3B}$=8, $O_{4B}$=0}, we also explored different options for determining the optimal LASSO α parameter based on the validation errors (Figure 5). In this study, we varied α over the range of $10^{-5} \leq α \leq 10^{-2}$. A more detailed summary of the results is shown in Table S2 in the Supplemental Materials section. In short, values of α = $10^{-2}$ and higher yields relatively high RMS errors in all of our validation tests, with RMS force errors (0.70 eV/Å in U and 0.33 eV/Å in H), lattice constants (-1.97% for α-U and -2.65% for α-UH$_3$), volume (-5.77% for α-U and -7.74% for α-UH$_3$), single uranium vacancy (-8.43%), single hydrogen interstitial (-100.37%). In contrast, the under-regularized value of α = $10^{-5}$ show considerable improvement with RMS force errors (0.52 eV/Å in U and 0.24 eV/Å in H), estimation of lattice parameters (-1.19% for α-U and -1.63% for α-UH$_3$), volume (-3.51% for α-U and -4.83% for α-UH$_3$), single uranium vacancy (20.17%), and single hydrogen interstitial (1.87%). However, as shown in Figure 5, the optimal balance of regularization and minimized training and validation occurs with α = $10^{-3}$. Here, compared to α = $10^{-5}$ model, we see slightly higher RMS force errors (0.52 vs 0.58 eV/Å in U and 0.24 vs 0.27 eV/Å in H), lattice (-1.19% vs -1.32% for α-U and -1.63% vs -1.9% for α-UH$_3$) and volume (-3.51% vs -3.90% for α-U and -4.83% vs -5.57% for α-UH$_3$) but significantly lower uranium vacancy formation energy errors (20.17% vs 12.53%) and similar hydrogen interstitial energy errors (1.87% vs 2.61%). Therefore, we chose to proceed with LASSO/LARS optimization with α = $10^{-3}$ as the best choice for our U-H model.

**1.4. Final hyperparameters.** Our final set of hyperparameters values includes $\{r_{min}^{H-H} = 0.50, r_{min}^{H-U} = 0.68, r_{min}^{U-U} = 1.07, r_{max}^{H-H} = 5.5, r_{max}^{H-U} = 5.5, r_{max}^{U-U} = 5.5\}$ and $\{O_{2B} = 16, O_{3B} = 8, O_{4B} = 0\}$, optimized with LASSO/LARS with a regularization of α = $10^{-3}$. This model yields the RMS force errors of 0.27 eV/Å in hydrogen and 0.58 eV/Å in uranium, the total energy of 0.24 eV/U atom, and the diagonal of the stress tensor of 0.77 GPa. For the remainder of our discussion, we will present results using this ChIMES model.

## B. Bulk structural properties

Using optimal parameters, we performed analysis of the pure α-U and α-UH$_3$, and β-UH$_3$. The ChIMES potential predicted lattice parameters of these reference structures, listed in the Table 1, agree quite well with DFT results[50] and experimental data[51,52]. The results for the α-U and α-UH$_3$ bulk properties indicate that the ChIMES yields lattice constants with errors of only ~1.19% and ~1.7% from the DFT and the experimentally determined values, respectively. Volumetric (per unit U atom) of the optimized unit cell and the bulk modulus parameters are slightly lower than DFT-determined results, with errors of 3.9% and 12.8% for α-U and 5.57 % and 21.4% for α-UH$_3$. In each case, the bulk modulus was estimated from energy vs. volume curve computed over a range of -13 to 24 GPa, followed by regression to a Birch-Murnaghan[53] model. We note that β-UH$_3$ data was not included in the training set in order to benchmark the transferability of the ChIMES potential. Our ChIMES potential is in excellent agreement with DFT-calculated and experimental values for β-UH$_3$, with relative errors of 2.1% and 3.0% for the lattice constant, and 6.1% and 8.7% for the unit cell volume, respectively. We compute an error of 0.9% for the bulk modulus relative to DFT. In addition, ChIMES yields the correct energetic ordering of the uranium hydride phases, with β-UH$_3$ predicted to be 0.03 eV/U atom lower in energy than α-UH$_3$, compared to the DFT computed result of 0.02 eV/U atom.

**Table 1:** Lattice parameters, unit cell volumes (V) per U atom, and bulk moduli (B) for α-U, α-UH$_3$, and β-UH$_3$ predicted by the ChIMES potential. Present results are compared with experimental and DFT calculations.

| Structure | Method | a (Å) | b (Å) | c (Å) | V (Å$^3$) | B (GPa) |
|---|---|---|---|---|---|---|
| α-U (ortho) | PBE | 2.81 | 5.87 | 4.92 | 20.29 | 148 |
| | ChIMES | 2.78 | 5.79 | 4.86 | 19.53 | 156 |
| | Exp | 2.84 | 5.87 | 4.94 | 20.59 | 125 |
| α-UH$_3$ (cubic) | PBE | 4.12 | 4.12 | 4.12 | 34.97 | 106 |
| | ChIMES | 4.05 | 4.05 | 4.05 | 33.22 | 135 |
| | Exp | 4.16 | 4.16 | 4.16 | 36 | --- |
| β-UH$_3$ (cubic) | PBE | 6.59 | 6.59 | 6.59 | 35.77 | 106 |
| | ChIMES | 6.45 | 6.45 | 6.45 | 33.54 | 115 |
| | Exp | 6.65 | 6.65 | 6.65 | 36.76 | --- |

## C. Additional point defect formation energies in α-U.

**1.1 Uranium vacancy energies as a function of system size.** In order to further validate our optimal ChIMES parameterization, we choose to compute the formation energies of vacancies in α-U as a function of supercell size, thus estimating the effect of uranium vacancy concentration. As shown in Figure 6, the vacancy formation energy shows some dependence on the uranium vacancy concentration in the bulk structure. The DFT estimated vacancy values were found to be 2.03 eV for (2x1x1), 1.93 eV for (2x2x1), 1.92 eV for (2x2x2), 1.90 eV for (3x2x2), 1.81 eV for (4x2x2), and 1.78 eV for (4x2x3) supercells. The 0.25 eV decrease in the vacancy formation energy with the increase of the system size from (2x1x1) to (4x2x3) indicates that there are strain-induced interactions around the defect at higher concentrations.

Our ChIMES calculated value in the dilute limit of 1.56 eV is 0.22 eV lower (~12% error) than our computed DFT value and about ~0.13-0.39 eV lower than other published DFT results of 1.95 eV (Taylor[54]), 1.69 eV (Wirth[55]), and 1.86 eV (Beeler[56]). We found that underestimation of the U vacancy energy was typical for all ChIMES models created in our study. We also observe a weaker dependence on vacancy formation energy as a function of defect concentration, where the curve from ChIMES is relatively flat compared to DFT. This yields somewhat larger error between the DFT and those predicted by ChIMES potentials at higher vacancy concentrations. This could be attributed to the lack of training data in our ChIMES model for multiple vacancies. We note that by positron annihilation experimentally determined vacancy formation energy[57] of 1.20±0.25 eV is relatively lower than DFT or ChIMES. In addition, DFT calculations could vary depending on choice of functional and dispersion interaction model.

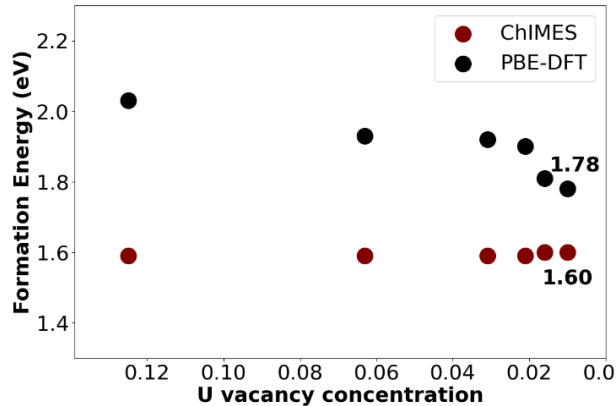

**Figure 6:** Uranium vacancy formation energy for α-U at various concentrations using ChIMES potential.

**1.2 Multiple hydrogen interstitial formation energies in α-U.** In this study, as shown in **Figure 7**, we have calculated the interstitial energies for (a) a low-energy interstitial hydrogen at the square-pyramidal (sq) position, a pair of hydrogen atoms at nearby sq sites in 2 different directions (b and c), and a hydrogen pair located ~5 Å apart (d). The square-pyramidal interstitial site occurs where the H atom is coordinated by five U atoms from the lattice (Figure 7a and Table 2). ChIMES predicts a formation energy for this site of 0.28 eV, which agrees within 0.01 of our result from DFT and is also in a good agreement with previously published results of 0.319 eV using 4x2x2 supercell (64 U atoms).[58] In addition, our ChIMES model predicts hydrogen interstitial formation energies that compare well to DFT for higher energy sites (not shown here), including the tetrahedral site with a value of 0.32 eV (~0.02 eV or 9.6% error) and the octahedral site with a value of 0.50 eV (~0.05 eV or 10.2% error). The tetrahedral and octahedral sites are relatively low energy and are likely thermally accessible under ambient conditions.

**Table 2.** Hydrogen interstitial formation energies (in eV) in α-U (4x2x3) supercell. The labeled systems are pictorially shown in Figure 7. The percent deviation relative to the DFT values are shown in parenthesis.

| System | DFT (eV) | ChIMES (eV) |
|---|---|---|
| sq (Figure 7a) | 0.27 | 0.28 (3.7%) |
| sq_sq_1 (Figure 7b) | 0.64 | 0.50 (-21.8%) |
| sq_sq_2 (Figure 7c) | 0.51 | 0.58 (13.7%) |
| sq_sq_3 (Figure 7d) | 0.52 | 0.65 (25.0%) |

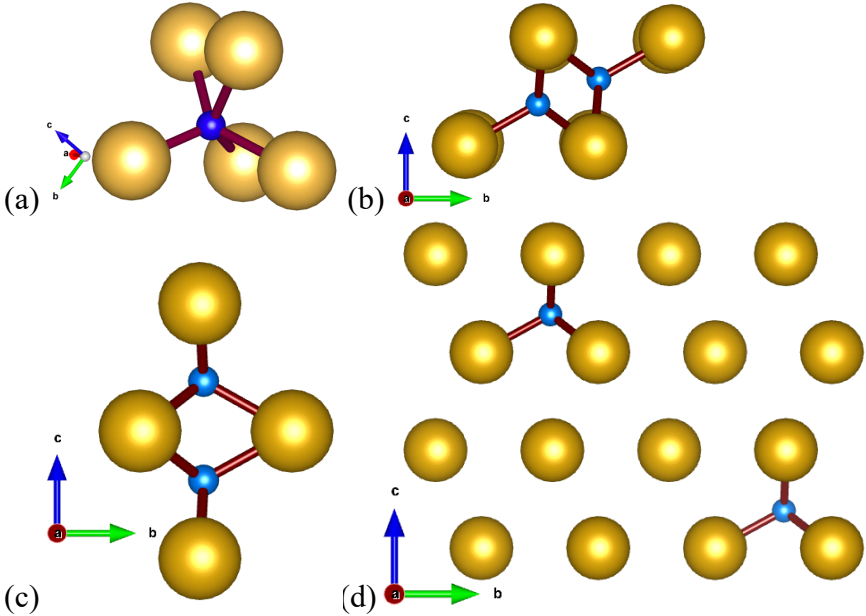

**Figure 7.** Hydrogen interstitial systems (a) square pyramidal site, (b) sq_sq_1, (c) sq_sq_2, and (d) sq_sq_3 systems.

We now examine our ChIMES model in terms of different double hydrogen interstitials in α-U with a 96 atom 4x2x3 supercell (Table 2 and Figure 7). For the short-range double interstitial system (two H interstitials in nearest-neighbor sites, labeled sq_sq_1) contains hydrogen atoms about 1.5 Å apart from each other. The ChIMES formation energy of 0.50 eV has an error of 0.14 eV, relative to DFT. On the other hand, the medium-range double interstitial system (H interstitials in next-nearest neighbor sites, labeled sq_sq_2) has an inter-hydrogen separation of 2.1 Å. Here, ChIMES yields a formation energy of 0.58 eV, with an error of 0.07 relative to DFT. Finally, the formation energy for the longer-range system (H interstitials several lattice spacings apart, labeled sq_sq_3) is also in relative agreement with the DFT value of 0.65 eV (a relative error of 0.17 eV). Overall, these results indicate that ChIMES can yield accurate physical quantities related to bulk hydrogen absorption within α-U lattices.

**1.3 Hydrogen vacancy in α-UH$_3$ as a function of concentration.** In addition to the uranium vacancy and hydrogen interstitials in α-U, we have also computed the hydrogen vacancy formation energy $E_{vac}$ in α-UH$_3$, defined as

$$E_{vac} = E_{def} + 1/2\, E_{H_2} - E_{perf},$$

(8)

where $E_{def}$ and $E_{perf}$ are the supercell energies for the defective and perfect systems, respectively, and $E_{H_2}$ is the energy of the isolated hydrogen molecule. The hydrogen vacancy formation energies were not part of our validation test and have been evaluated at various concentrations. Here, we begin with α-UH₃ supercell size of 3x3x3 (54U + 162H atoms), sequentially remove random hydrogen(s), and optimize the atomic positions. As shown in Figure **8**, the defect formation energy is relatively constant over the concentration range probed in our analysis. The calculated DFT $E_{vac}$ value for the bulk α-UH₃ phase is 0.90 eV, with a ChIMES predicted value of 0.85 eV However, we observe that both ChIMES and DFT results are relatively flat as a function of hydrogen vacancy concentration.

All of the results presented in this subsection indicate that ChIMES exhibits a high degree of accuracy for different bulk uranium and UH₃ properties. In particular, our model yields accurate results for a number of validation that were not included in our training set. This includes the relative energetic ordering of the two UH₃ phases, uranium vacancy and hydrogen interstitial formation energies in α-U as under varying system sizes and concentrations, and hydrogen vacancy formation over a range of concentrations in α-UH₃. These all indicate that ChIMES can yield close to DFT accuracy for the energetics of U/H-containing systems under a variety of conditions, allowing us to use our model for kinetic and molecular dynamics calculations relevant to the hydriding process.

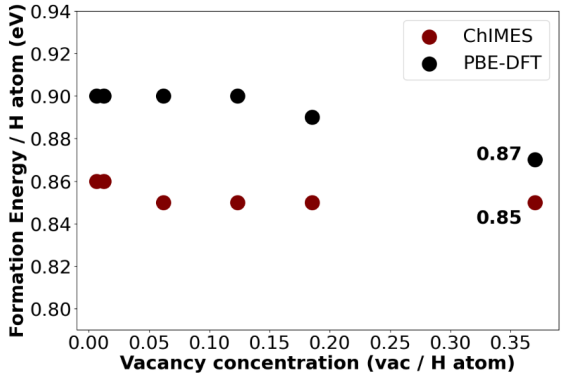

**Figure 8**. Hydrogen vacancy formation energy for α-UH₃ at various concentrations.

### D. Hydrogen hopping barriers in α-U bulk.

We now compute kinetic parameters for hydrogen diffusion between square pyramidal sites within bulk α-U. The atomic hydrogen diffusion minimum energy pathways (MEP) were calculated via the climbing image nudged elastic band (NEB) method.[59] Here, a chain of 3-5 linearly interpolated images along an initial pathway between initial and final sq absorption sites was relaxed to determine the MEP and its corresponding saddle point. Calculations were performed until the maximum residual forces on each atom was converged to less than 0.01 eV/Å. The transition state was confirmed by presence of one imaginary vibrational frequency.

Two main sq-sq pathways were identified: (1) along the <011> lattice direction, and (2) along the <001> direction, as illustrated in Figure 9. Results of this study (Figure 10) show that diffusion barrier and the jump distance depend on the pathway direction. For pathway 1, the hopping distance between sites for hydrogen migration is 1.5 Å and the DFT barrier height is 0.14 eV. For pathway 2, the hopping distance is 2.1 Å and the DFT estimated barrier is 0.38 eV. For comparison, ChIMES yields a calculated barrier is 0.09 eV (0.05 eV relative error) for pathway 1 and 0.39 eV (0.01 eV residual error) for pathway 2, indicative of a high degree of accuracy. These barrier values are in close proximity with the experimental result for bulk hydrogen diffusion of 0.280 eV,[60] which likely represents an average quantity for an imperfect crystalline system.

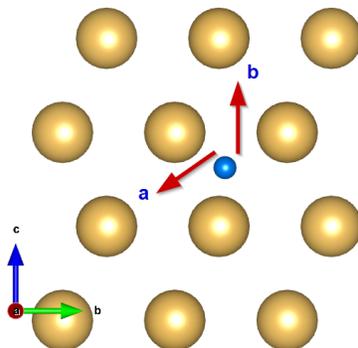

**Figure 9:** Pictorial representation of two potential pathways: (a) pathway along <011> direction, (b) pathway along <001>.

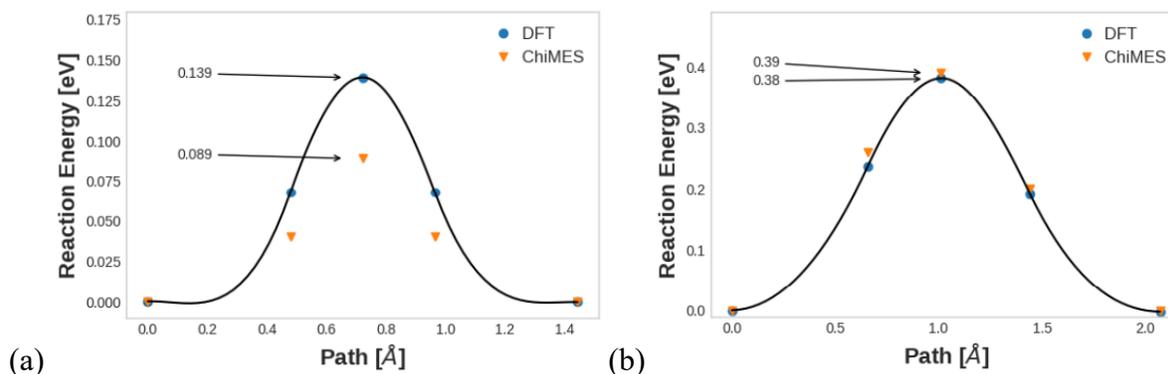

**Figure 10:** NEB predicted barrier for hydrogen diffusion from sqpy to another nearby sqpy interstitial site along (a) <011> direction and (b) <001>.

**E. Molecular dynamics validation.**

We now evaluate the reliability of our ChIMES model for molecular dynamic (MD) simulations by comparing energies with DFT along a computed trajectory. This was performed by first computing a short NVT MD simulation with ChIMES of 5x2x3 α-U supercell (120 U atoms) at 400 K. The trajectory was computed for ~20 picosecond with a time step of 4.0 femtoseconds. We then extracted images after every 100th MD step and calculated single point energies using DFT in order to determine errors in the resulting energies. Figure 11 shows the ChIMES predicted energies of these structures along the MD trajectory in comparison to their DFT reference values. We observe accurate system energies from ChIMES, with slight overestimations of the DFT values with relative errors up to 0.05 eV/atom. Similar MD studies were performed for α-$UH_3$ (54U + 162H) and β-$UH_3$ (64U + 192H) systems, shown in Figure 12. In each of these systems, we computed trajectories for ~5 ps with a time step of 0.2 fs and collecting images every 1000 steps. For both of these systems, we observe errors in the system energies 0.10 eV/U atom in α-$UH_3$ and below 0.15 eV/U atom in β-$UH_3$.

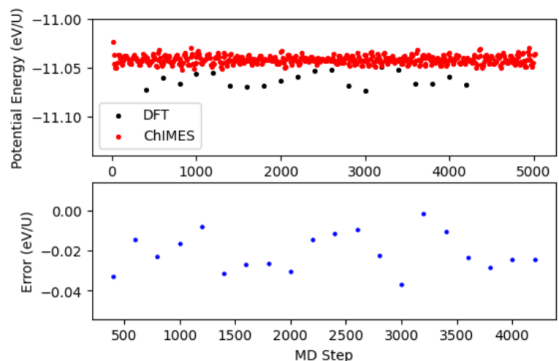

**Figure 11**: Comparison of energy predicted by the ChIMES and DFT along the same MD trajectory in α-U.

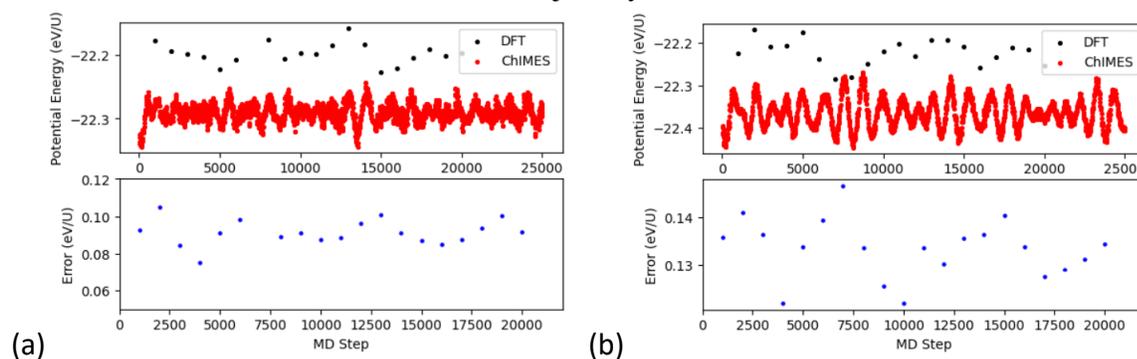

(a)  (b)

**Figure 12**. Comparison of energy predicted by the ChIMES and DFT along the same MD trajectory in (a) α-UH$_3$ and (b) β-UH$_3$.

We also have used MD simulations with our ChIMES potential to investigate the frequency of sites the hydrogen interstitial hopping in bulk α-U. For this study, we have performed ten 10 ps NVT MD simulations over a range of temperatures (100-1000K) of one atomic hydrogen atom in a large α-U supercell (840U + 1H atoms) with a time step of 0.2 fs. Hydrogen location was monitored based on the number of U atoms surrounding interstitial H atom a given radius. In this case, a value of 2.85 Å centered at each site was chosen based on the largest possible U-H distance, which is the out of plane distance along the stretched axis in the octahedral site (~2.65 A°). An additional 0.2 Å was added in order to compensate for thermal fluctuations of lattice sites during the MD simulation. Hydrogen locations were determined at each step and residency time was monitored during the molecular dynamics simulation.

As shown in Figure 13, hydrogen predominantly occupies the square-pyramidal interstitial site for all temperatures of our study with residence of over 50%. The square-pyramidal is the most stable interstitial site in pure uranium bulk, which is confirmed by our ChIMES potential and DFT studies and has the shortest average residency time of 1.7 fs. This short residency time is commensurate with the relatively low hopping barriers computed for both <011> and <001> diffusion pathways. As temperature increases, the fraction of the square-pyramidal site decreases from 0.67 (100 K) to 0.46 (1000 K), as the other sites – tetrahedral (+0.05 eV greater absorption energy) and octahedral (+0.22 eV greater than square-pyramidal) – become more energetically accessible. The second most stable interstitial tetrahedral site shows occupancy fraction of ~0.30, which remains relatively constant for all range of temperatures, with an average residency time of 4.44 fs. On the other hand, we observe the first occurrence of the octahedral site at a temperature of 300 K. As the system temperature increases, the frequency of occurrence of the octahedral site also increases, apparently at the expense of the square-pyramidal site. The octahedral was found to have the longest residency time of 8.39 fs. Overall, all residency times computed here are exceedingly short at ambient conditions. This indicates that hydrogen are highly mobile in the pure metal lattice, which likely has ramifications for hydriding initiation.

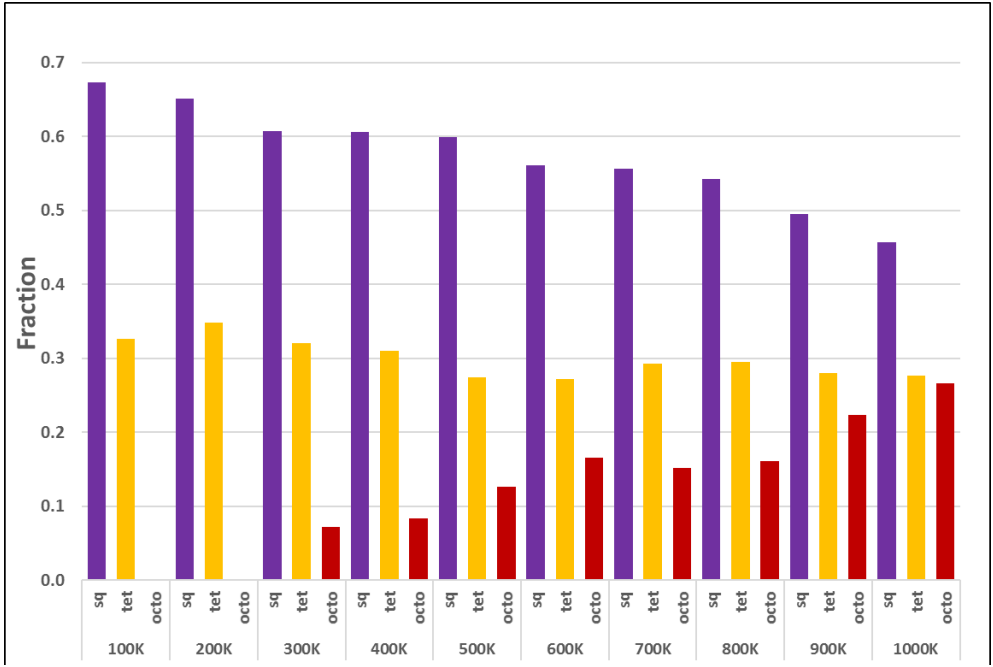

**Figure 13.** Interstitial site type analysis.

### F. Hydrogen diffusivity as a function of temperature.

Finally, we have computed hydrogen diffusion coefficients in pristine α-U as a function of temperature from MD simulations. During the constant temperature MD simulation, the mean square displacement (MSD) of hydrogen atoms was calculated and the diffusion coefficient was determined from the standard Einstein-Smoluchowski relation. Each MD simulation (840 U + 20 H atoms with supercell dimensions of 19.19 x 29.23 x 29.24 Å, similar to α-U at ambient density) was run for 10 ps with time step of 0.2 fs. The MSD was averaged over a time interval of at least 5ps for each calculation. In addition, the diffusivities were then estimated by linear regression of the MSDs over a temperature range of 300 K to 1000 K and then fit to the Arrhenius equation:

$$D(T) = D_0 \exp\left(-\frac{E_a}{k_B T}\right).$$

(9)

Here, $D_0$ is a prefactor, $k_B$ is the Boltzmann constant and $E_a$ is a total hydrogen diffusion activation energy. The Arrhenius plot of hydrogen diffusivity as a function of temperature is shown in Figure 14. Our regression analysis yields a value of $D_0$ of $2.73 \times 10^{-3}$ cm$^2$/s, which is in very good agreement with the experimentally determined result of $1.43 \times 10^{-3}$ cm$^2$/s.[60]

However, ChIMES yields an overall diffusion barrier of 0.13 eV, which a bit lower than experimental value of 0.28 eV (Mallett[60]). This could be attributed to the environmental conditions surrounding the metal during experimentation, such as presence of the surface cracks, grain-boundaries, and the passive layer of oxides, oxycarbides, and water, while our computational simulation currently probes a defect-free crystal.

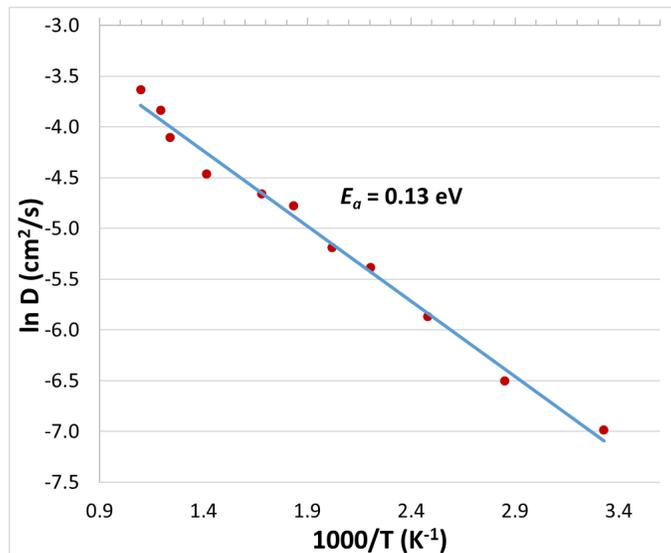

**Figure 14**: Arrhenius plot of hydrogen diffusivity at temperatures ranging from 300 to 1000 K.

## CONCLUSION

In this work, we have utilized an optimzation workflow to create a ChIMES reactive force field for a variety of uranium-hydrogen containing systems. Optimal ChIMES parameters were determined extremely rapidly in a semi-automated approach by varying the minimum and maximum pairwise interaction radius, the polynomial order, and the type of the regularization. Overall, we find that our ChIMES model yields comparable accuracy to DFT for the U-H containing systems studied here. This includes thermodynamic data, such as the bulk structural parameters of $\alpha$-U, $\alpha$-UH$_3$, and $\beta$-UH$_3$, the relative energetic ordering of UH$_3$ phases, and the bulk moduli of these three materials. Our model also yields accurate defect formation energies in both $\alpha$-U and $\alpha$-UH$_3$ over a range of defect concentrations. Finally, ChIMES yields accurate kinetic data for hydrogen interstitial hops in an a-U lattice as well as bulk diffusivity over a broad temperature range. The linear scaling of ChIMES and computational efficiency relative to standard DFT allows for its use in large-scale MD simulation, which could include amorphous grain boundaries and hydride phase nucleation and growth. Future efforts will involve simulation of these hydriding phenomena in uranium, including formation of the hydride metal within the bulk around defect sites. Our efforts will allow us to make more direct contact with experiments, where atomic-level simulations can be valuable tools for elucidation and interpretation of results.

# SUPPLEMENTAL MATERIALS

**Table S1:** Validation Results for polynomial structure a) RMS, b) Validation tests, and c) Validation test (including 4B interactions). Polynomial orders (column 1) are listed in order of two-body, three-body, and four-body. The headings 'rms H' and 'rms U' corresponds to the root mean square errors in the forces on U and H atoms.

(a)

| Polynomial Type | rms H (ev/Å) | rms U (ev/Å) | rms E (ev/atom) | rms P (GPa) | Polynomial Type | rms H (ev/Å) | rms U (ev/Å) | rms E (ev/atom) | rms P (GPa) |
|---|---|---|---|---|---|---|---|---|---|
| 2 2 0 | 2.024 | 20.521 | 3.660 | 26.16 | 10 8 0 | 0.31 | 0.871 | 0.251 | 1.481 |
| 2 4 0 | 0.843 | 10.984 | 3.412 | 50.497 | 10 10 0 | 0.239 | 0.539 | 0.263 | 0.867 |
| 2 6 0 | 0.634 | 2.778 | 1.633 | 17.645 | 10 12 0 | 0.253 | 0.506 | 0.346 | 1.273 |
| 2 8 0 | 0.569 | 2.096 | 1.670 | 17.427 | 12 2 0 | 0.713 | 16.029 | 1.246 | 4.287 |
| 2 10 0 | 0.625 | 1.797 | 1.579 | 16.331 | 12 4 0 | 0.994 | 8.093 | 2.019 | 14.789 |
| 2 12 0 | 0.488 | 1.577 | 1.446 | 14.988 | 12 6 0 | 0.488 | 1.834 | 0.371 | 2.662 |
| 2 14 0 | 0.441 | 1.462 | 1.443 | 14.328 | 12 8 0 | 0.268 | 0.623 | 0.261 | 1.137 |
| 4 2 0 | 1.646 | 19.722 | 9.886 | 114.472 | 12 10 0 | 0.223 | 0.426 | 0.302 | 0.927 |
| 4 4 0 | 0.909 | 9.77 | 6.729 | 53.609 | 12 12 0 | 0.202 | 0.359 | 0.385 | 0.729 |
| 4 6 0 | 0.609 | 2.681 | 0.961 | 10.381 | 14 2 0 | 0.709 | 16.026 | 1.223 | 4.095 |
| 4 8 0 | 0.538 | 1.974 | 1.216 | 12.211 | 14 4 0 | 0.991 | 8.092 | 2.018 | 14.774 |
| 4 10 0 | 0.589 | 1.654 | 1.022 | 10.186 | 14 6 0 | 0.497 | 1.818 | 0.403 | 2.921 |
| 4 12 0 | 0.45 | 1.404 | 0.797 | 7.843 | 14 8 0 | 0.266 | 0.573 | 0.247 | 0.769 |
| 4 14 0 | 0.399 | 1.194 | 0.771 | 6.641 | 14 10 0 | 0.22 | 0.403 | 0.301 | 0.727 |
| 6 2 0 | 1.157 | 18.839 | 6.783 | 75.467 | 14 12 0 | 0.202 | 0.368 | 0.381 | 0.705 |
| 6 4 0 | 0.983 | 9.339 | 5.912 | 44.722 | 16 2 0 | 0.701 | 16.025 | 1.148 | 3.808 |
| 6 6 0 | 0.546 | 2.169 | 0.224 | 2.531 | 16 4 0 | 0.987 | 8.062 | 2.039 | 14.543 |
| 6 8 0 | 0.431 | 1.544 | 0.667 | 6.48 | 16 6 0 | 0.497 | 1.815 | 0.406 | 2.97 |
| 6 10 0 | 0.454 | 1.153 | 0.487 | 4.56 | 16 8 0 | 0.266 | 0.58 | 0.243 | 0.771 |
| 6 12 0 | 0.335 | 0.923 | 0.355 | 2.924 | 16 8 1 | 0.266 | 0.58 | 0.243 | 0.771 |
| 6 14 0 | 0.287 | 0.713 | 0.495 | 2.094 | 16 8 2 | 0.263 | 0.56 | 0.238 | 0.738 |
| 8 2 0 | 0.748 | 16.581 | 2.403 | 22.207 | 16 8 3 | 0.254 | 0.532 | 0.216 | 0.789 |
| 8 4 0 | 0.999 | 8.441 | 3.365 | 23.168 | 16 8 4 | 0.239 | 0.512 | 0.191 | 0.877 |
| 8 6 0 | 0.489 | 1.971 | 0.283 | 1.962 | 16 10 0 | 0.223 | 0.412 | 0.296 | 0.754 |
| 8 8 0 | 0.338 | 1.037 | 0.314 | 3.165 | 16 12 0 | 0.201 | 0.356 | 0.400 | 0.711 |
| 8 10 0 | 0.265 | 0.635 | 0.243 | 1.691 | 18 2 0 | 0.697 | 16.024 | 1.127 | 3.638 |
| 8 12 0 | 0.233 | 0.474 | 0.276 | 1.182 | 18 4 0 | 0.997 | 8.048 | 2.034 | 14.614 |
| 8 14 0 | 0.203 | 0.369 | 0.481 | 1 | 18 6 0 | 0.501 | 1.806 | 0.411 | 3.007 |
| 10 2 0 | 0.715 | 16.129 | 1.186 | 5.655 | 18 8 0 | 0.266 | 0.571 | 0.248 | 0.73 |
| 10 4 0 | 0.985 | 8.153 | 2.222 | 15.027 | 18 10 0 | 0.219 | 0.401 | 0.304 | 0.709 |
| 10 6 0 | 0.475 | 1.938 | 0.323 | 2.371 | 18 12 0 | 0.197 | 0.342 | 0.421 | 0.696 |

(b)

| Poly Type | aU_lata | aU_latb | aU_latc | aU_V | aUH3_lata | aUH3_V | DEF Vac_U | DEF Int1H |
|---|---|---|---|---|---|---|---|---|
| 2 2 0 | 2.542 (-9.68%) | 5.3 (-9.67%) | 4.448 (-9.68%) | 14.98 (-26.29%) | 3.749 (-9.01%) | 26.34 (-24.67%) | -178.505 (10128.) | -157.872 (59007.46%) |
| 2 4 0 | 2.542 (-9.68%) | 5.3 (-9.67%) | 4.448 (-9.68%) | 14.98 (-26.29%) | 3.455 (-16.15%) | 20.62 (-41.03%) | 127.921 (7086.57) | -34.113 (12828.73%) |
| 2 6 0 | 2.417 (-14.11%) | 5.04 (-14.1%) | 4.229 (-14.11%) | 12.88 (-36.63%) | 3.58 (-13.1%) | 22.95 (-34.37%) | 10.101 (467.47%) | -31.461 (11839.18%) |
| 2 8 0 | 2.388 (-15.15%) | 4.979 (-15.14%) | 4.178 (-15.15%) | 12.42 (-38.89%) | 3.575 (-13.24%) | 22.84 (-34.68%) | -18.176 (1121.12) | -18.679 (7069.78%) |
| 2 10 0 | 2.475 (-12.04%) | 5.161 (-12.03%) | 4.331 (-12.04%) | 13.83 (-31.95%) | 3.658 (-11.21%) | 24.48 (-29.99%) | 5.612 (215.28%) | 1.774 (561.94%) |
| 2 12 0 | 2.463 (-12.47%) | 5.136 (-12.46%) | 4.31 (-12.47%) | 13.63 (-32.93%) | 3.611 (-12.36%) | 23.53 (-32.71%) | 3.435 (92.98%) | 3.721 (1288.43%) |
| 2 14 0 | 2.505 (-10.97%) | 5.224 (-10.95%) | 4.384 (-10.97%) | 14.35 (-29.39%) | 3.59 (-12.87%) | 23.13 (-33.85%) | -17.21 (1066.859) | -20.405 (7713.81%) |
| 4 2 0 | 2.307 (-18.01%) | 4.811 (-18.0%) | 4.037 (-18.01%) | 11.2 (-44.89%) | 3.442 (-16.45%) | 20.39 (-41.69%) | 60.935 (3323.319) | 76.408 (28410.45%) |
| 4 4 0 | 6.092 (116.5%) | 12.704 (116.52%) | 10.66 (116.5%) | 206.26 (914.88%) | 6.435 (56.19%) | 133.25 (281.07%) | -0.0 (100.0%) | -17.683 (6698.13%) |
| 4 6 0 | 2.37 (-15.79%) | 4.941 (-15.78%) | 4.146 (-15.79%) | 12.14 (-40.27%) | 3.752 (-8.92%) | 26.42 (-24.44%) | 10.299 (478.6%) | -20.119 (7607.09%) |
| 4 8 0 | 2.592 (-7.88%) | 5.405 (-7.87%) | 4.536 (-7.88%) | 15.89 (-21.81%) | 3.589 (-12.88%) | 23.12 (-33.88%) | -80.362 (4614.72) | -93.514 (34993.28%) |
| 4 10 0 | 2.579 (-8.35%) | 5.378 (-8.34%) | 4.513 (-8.35%) | 15.65 (-23.0%) | 3.757 (-8.82%) | 26.5 (-24.21%) | -22.233 (1349.04) | -43.832 (16455.22%) |
| 4 12 0 | 2.645 (-6.01%) | 5.515 (-5.99%) | 4.628 (-6.01%) | 16.88 (-16.94%) | 3.697 (-10.28%) | 25.26 (-27.76%) | 1.425 (19.94%) | -0.883 (429.48%) |
| 4 14 0 | 2.657 (-5.57%) | 5.541 (-5.55%) | 4.65 (-5.57%) | 17.12 (-15.76%) | 3.825 (-7.15%) | 27.99 (-19.95%) | 0.714 (59.89%) | -0.433 (261.57%) |
| 6 2 0 | 2.241 (-20.37%) | 4.672 (-20.36%) | 3.921 (-20.37%) | 10.26 (-49.52%) | 3.383 (-17.89%) | 19.36 (-44.63%) | 59.738 (3256.079) | -105.326 (39400.75%) |
| 6 4 0 | 6.092 (116.5%) | 12.704 (116.52%) | 10.66 (116.5%) | 206.26 (914.88%) | 6.423 (55.89%) | 132.46 (278.81%) | -6.884 (486.74%) | -14.201 (5398.88%) |
| 6 6 0 | 2.673 (-5.0%) | 5.575 (-4.98%) | 4.678 (-5.0%) | 17.43 (-14.24%) | 3.99 (-3.17%) | 31.75 (-9.2%) | 1.304 (26.74%) | -1.353 (604.85%) |
| 6 8 0 | 2.654 (-5.7%) | 5.533 (-5.69%) | 4.643 (-5.7%) | 17.04 (-16.1%) | 3.911 (-5.06%) | 29.92 (-14.43%) | 4.388 (146.52%) | 1.479 (451.87%) |
| 6 10 0 | 2.682 (-4.69%) | 5.593 (-4.68%) | 4.693 (-4.69%) | 17.6 (-13.4%) | 3.887 (-5.65%) | 29.37 (-16.01%) | 4.097 (130.17%) | 0.087 (67.54%) |
| 6 12 0 | 2.711 (-3.67%) | 5.652 (-3.66%) | 4.743 (-3.67%) | 18.17 (-10.6%) | 3.931 (-4.58%) | 30.38 (-13.12%) | 3.516 (97.53%) | 0.571 (113.06%) |
| 6 14 0 | 2.741 (-2.61%) | 5.715 (-2.6%) | 4.795 (-2.61%) | 18.78 (-7.59%) | 3.982 (-3.36%) | 31.56 (-9.74%) | 2.317 (30.17%) | 0.175 (34.7%) |
| 8 2 0 | 2.36 (-16.13%) | 4.921 (-16.12%) | 4.13 (-16.13%) | 11.99 (-41.0%) | 3.435 (-16.62%) | 20.27 (-42.03%) | 11.412 (541.12%) | -4.462 (1764.93%) |
| 8 4 0 | 6.092 (116.5%) | 12.704 (116.52%) | 10.66 (116.5%) | 206.26 (914.88%) | 6.38 (54.85%) | 129.82 (271.26%) | -16.489 (1026.35) | 7.126 (2558.96%) |
| 8 6 0 | 2.68 (-4.76%) | 5.588 (-4.75%) | 4.689 (-4.76%) | 17.56 (-13.6%) | 4.089 (-0.76%) | 34.17 (-2.28%) | 3.316 (86.29%) | 0.515 (92.16%) |
| 8 8 0 | 2.697 (-4.16%) | 5.624 (-4.15%) | 4.719 (-4.16%) | 17.89 (-11.97%) | 3.973 (-3.57%) | 31.35 (-10.34%) | 3.005 (68.82%) | -0.087 (132.46%) |
| 8 10 0 | 2.745 (-2.44%) | 5.725 (-2.43%) | 4.804 (-2.44%) | 18.87 (-7.15%) | 3.983 (-3.31%) | 31.61 (-9.6%) | 1.986 (11.57%) | 0.392 (46.27%) |
| 8 12 0 | 2.76 (-1.9%) | 5.756 (-1.89%) | 4.83 (-1.9%) | 19.19 (-5.58%) | 3.997 (-2.98%) | 31.93 (-8.69%) | 1.72 (3.37%) | -0.071 (126.49%) |
| 8 14 0 | 2.758 (-1.99%) | 5.751 (-1.98%) | 4.826 (-1.99%) | 19.14 (-5.82%) | 4.013 (-2.59%) | 32.32 (-7.57%) | 1.632 (8.31%) | -0.432 (261.19%) |
| 10 2 0 | 2.931 (4.17%) | 6.112 (4.18%) | 5.129 (4.17%) | 22.97 (13.02%) | 3.551 (-13.81%) | 22.39 (-35.97%) | 5.988 (236.4%) | -4.497 (1777.99%) |
| 10 4 0 | 3.551 (26.17%) | 7.404 (26.19%) | 6.213 (26.17%) | 40.83 (100.9%) | 6.317 (53.33%) | 126.05 (260.48%) | -5.207 (392.53%) | -1.808 (774.63%) |
| 10 6 0 | 6.092 (116.5%) | 12.704 (116.52%) | 10.66 (116.5%) | 206.26 (914.88%) | 4.134 (0.34%) | 35.32 (1.01%) | -272.121 (15387.) | -267.005 (99728.73%) |
| 10 8 0 | 2.749 (-2.33%) | 5.731 (-2.31%) | 4.809 (-2.33%) | 18.94 (-6.81%) | 4.005 (-2.78%) | 32.13 (-8.11%) | 2.026 (13.82%) | -0.01 (103.73%) |
| 10 10 0 | 2.774 (-1.41%) | 5.785 (-1.4%) | 4.854 (-1.41%) | 19.48 (-4.15%) | 4.032 (-2.14%) | 32.77 (-6.28%) | 1.496 (15.96%) | 0.216 (19.4%) |
| 10 12 0 | 2.752 (-2.21%) | 5.738 (-2.19%) | 4.815 (-2.21%) | 19.01 (-6.46%) | 3.982 (-3.34%) | 31.58 (-9.69%) | 1.694 (4.83%) | -0.358 (233.58%) |
| 12 2 0 | 2.931 (4.17%) | 6.112 (4.18%) | 5.129 (4.17%) | 22.97 (13.02%) | 3.438 (-16.55%) | 20.32 (-41.89%) | 8.92 (401.12%) | -2.907 (1184.7%) |
| 12 4 0 | 3.626 (28.86%) | 7.561 (28.88%) | 6.345 (28.86%) | 43.49 (113.99%) | 6.304 (53.01%) | 125.27 (258.25%) | -3.087 (273.43%) | -1.241 (563.06%) |
| 12 6 0 | 6.092 (116.5%) | 12.704 (116.52%) | 10.66 (116.5%) | 206.26 (914.88%) | 4.158 (0.92%) | 35.94 (2.78%) | -296.043 (16731.) | -277.507 (103647.39%) |
| 12 8 0 | 2.749 (-2.32%) | 5.732 (-2.31%) | 4.81 (-2.32%) | 18.94 (-6.81%) | 4.032 (-2.13%) | 32.78 (-6.26%) | 1.796 (0.9%) | 0.223 (16.79%) |
| 12 10 0 | 2.767 (-1.67%) | 5.77 (-1.66%) | 4.842 (-1.67%) | 19.33 (-4.89%) | 4.039 (-1.98%) | 32.93 (-5.83%) | 1.401 (21.29%) | 0.187 (30.22%) |
| 12 12 0 | 2.779 (-1.23%) | 5.796 (-1.21%) | 4.864 (-1.23%) | 19.59 (-3.61%) | 4.037 (-2.01%) | 32.9 (-5.91%) | 1.303 (26.8%) | -0.07 (126.12%) |
| 14 2 0 | 2.931 (4.17%) | 6.112 (4.18%) | 5.129 (4.17%) | 22.97 (13.02%) | 3.33 (-19.19%) | 18.46 (-47.21%) | 9.002 (405.73%) | -2.823 (1153.36%) |
| 14 4 0 | 3.622 (28.72%) | 7.553 (28.74%) | 6.338 (28.72%) | 43.35 (113.3%) | 6.309 (53.13%) | 125.57 (259.11%) | -3.142 (276.52%) | -1.075 (501.12%) |
| 14 6 0 | 6.092 (116.5%) | 12.704 (116.52%) | 10.66 (116.5%) | 206.26 (914.88%) | 4.173 (1.28%) | 36.33 (3.9%) | 0.34 (80.9%) | 0.11 (58.96%) |
| 14 8 0 | 2.773 (-1.47%) | 5.782 (-1.45%) | 4.852 (-1.47%) | 19.44 (-4.35%) | 4.043 (-1.88%) | 33.04 (-5.51%) | 1.568 (11.91%) | 0.245 (8.58%) |
| 14 10 0 | 2.78 (-1.22%) | 5.796 (-1.21%) | 4.864 (-1.22%) | 19.59 (-3.61%) | 4.045 (-1.82%) | 33.1 (-5.34%) | 1.315 (26.12%) | 0.173 (35.45%) |
| 14 12 0 | 2.782 (-1.13%) | 5.802 (-1.11%) | 4.869 (-1.13%) | 19.65 (-3.31%) | 4.035 (-2.06%) | 32.85 (-6.05%) | 1.306 (26.63%) | -0.005 (101.87%) |
| 16 2 0 | 2.931 (4.17%) | 6.112 (4.18%) | 5.129 (4.17%) | 22.97 (13.02%) | 3.544 (-13.98%) | 22.26 (-36.34%) | 9.072 (409.66%) | -2.488 (1028.36%) |
| 16 4 0 | 3.548 (26.07%) | 7.398 (26.09%) | 6.208 (26.07%) | 40.73 (100.41%) | 6.311 (53.18%) | 125.69 (259.45%) | -3.589 (301.63%) | -1.324 (594.03%) |
| 16 6 0 | 6.092 (116.5%) | 12.704 (116.52%) | 10.66 (116.5%) | 206.26 (914.88%) | 4.184 (1.55%) | 36.62 (4.73%) | -3.48 (295.51%) | 0.927 (245.9%) |
| 16 8 0 | 2.777 (-1.32%) | 5.79 (-1.3%) | 4.859 (-1.32%) | 19.53 (-3.9%) | 4.042 (-1.9%) | 33.02 (-5.57%) | 1.557 (12.53%) | 0.275 (2.61%) |
| 16 10 0 | 2.776 (-1.35%) | 5.788 (-1.34%) | 4.857 (-1.35%) | 19.51 (-4.0%) | 4.043 (-1.86%) | 33.05 (-5.48%) | 1.249 (29.83%) | 0.082 (69.4%) |
| 16 12 0 | 2.779 (-1.23%) | 5.796 (-1.22%) | 4.864 (-1.23%) | 19.59 (-3.61%) | 4.036 (-2.04%) | 32.87 (-6.0%) | 1.37 (23.03%) | -0.032 (111.94%) |
| 18 2 0 | 2.931 (4.17%) | 6.112 (4.18%) | 5.129 (4.17%) | 22.97 (13.02%) | 3.59 (-12.86%) | 23.14 (-33.82%) | 8.96 (403.37%) | -2.276 (949.25%) |
| 18 4 0 | 3.554 (26.3%) | 7.411 (26.32%) | 6.219 (26.3%) | 40.95 (101.49%) | 6.31 (53.17%) | 125.64 (259.31%) | -3.259 (283.09%) | -1.383 (616.04%) |
| 18 6 0 | 6.092 (116.5%) | 12.704 (116.52%) | 10.66 (116.5%) | 206.26 (914.88%) | 4.166 (1.12%) | 36.15 (3.38%) | -3.746 (310.45%) | 0.903 (236.94%) |
| 18 8 0 | 2.775 (-1.4%) | 5.785 (-1.39%) | 4.855 (-1.4%) | 19.48 (-4.15%) | 4.045 (-1.83%) | 33.08 (-5.4%) | 1.514 (14.94%) | 0.264 (1.49%) |
| 18 10 0 | 2.779 (-1.24%) | 5.795 (-1.23%) | 4.863 (-1.24%) | 19.58 (-3.66%) | 4.047 (-1.77%) | 33.14 (-5.23%) | 1.3 (26.97%) | 0.15 (44.03%) |
| 18 12 0 | 2.781 (-1.16%) | 5.8 (-1.14%) | 4.867 (-1.15%) | 19.63 (-3.41%) | 4.04 (-1.94%) | 32.97 (-5.71%) | 1.393 (21.74%) | 0.041 (84.7%) |

(c)

| Poly Type | aU_lata | aU_latb | aU_latc | aU_V | aUH3_lata | aUH3_V | DEF Vac_U | DEF Int1H |
|---|---|---|---|---|---|---|---|---|
| 16 8 0 | 2.777 (-1.32%) | 5.79 (-1.3%) | 4.859 (-1.32%) | 19.53 (-3.9%) | 4.042 (-1.9%) | 33.02 (-5.57%) | 1.557 (12.53%) | 0.275 (2.61%) |
| 16 8 1 | 2.777 (-1.32%) | 5.79 (-1.3%) | 4.859 (-1.32%) | 19.53 (-3.9%) | 4.042 (-1.9%) | 33.02 (-5.57%) | 1.557 (12.53%) | 0.275 (2.61%) |
| 16 8 2 | 2.776 (-1.34%) | 5.789 (-1.32%) | 4.858 (-1.34%) | 19.52 (-3.95%) | 4.074 (-1.12%) | 33.8 (-3.34%) | 1.599 (10.17%) | 0.244 (8.96%) |
| 16 8 3 | 2.765 (-1.75%) | 5.765 (-1.74%) | 4.838 (-1.75%) | 19.28 (-5.13%) | 4.079 (-1.01%) | 33.92 (-2.99%) | 1.528 (12.54%) | 0.374 (39.55%) |
| 16 8 4 | 2.77 (-1.56%) | 5.776 (-1.55%) | 4.847 (-1.56%) | 19.39 (-4.59%) | 4.082 (-0.92%) | 34.01 (-2.74%) | 1.514 (14.94%) | 0.384 (43.28%) |

**Table S2**: Summary of validation tests using different values of LASSO/LARS parameter α:
a) RMS, b) Validation tests.

a)

| α value | rms H (eV/Å) | rms U (eV/Å) | rms Energy (eV) | rms P (GPa) |
|---|---|---|---|---|
| 1.00E-05 | 0.238 | 0.518 | 0.249 | 0.674 |
| 1.00E-04 | 0.243 | 0.532 | 0.250 | 0.684 |
| 1.00E-03 | 0.266 | 0.58 | 0.243 | 0.771 |
| 1.00E-02 | 0.326 | 0.702 | 0.253 | 1.234 |

b)

| α value | α-U | | | | | | α-UH$_3$ | |
|---|---|---|---|---|---|---|---|---|
| | lat_a (Å) | lat_b (Å) | lat_c (Å) | Vol. (Å$^3$) | Vac 1U (eV) | Int 1H (eV) | lat (Å) | Vol. (Å$^3$) |
| 1.00E-05 | 2.78 (-1.19%) | 5.798 (-1.18%) | 4.865 (-1.19%) | 19.61 (-3.51%) | 1.421 (20.17%) | 0.263 (1.87%) | 4.053 (-1.63%) | 33.28 (-4.83%) |
| 1.00E-04 | 2.781 (-1.19%) | 5.798 (-1.17%) | 4.866 (-1.18%) | 19.61 (-3.51%) | 1.409 (20.84%) | 0.223 (16.79%) | 4.051 (-1.67%) | 33.24 (-4.94%) |
| 1.00E-03 | 2.777 (-1.32%) | 5.79 (-1.3%) | 4.859 (-1.32%) | 19.53 (-3.9%) | 1.557 (12.53%) | 0.275 (2.61%) | 4.042 (-1.9%) | 33.02 (-5.57%) |
| 1.00E-02 | 2.759 (-1.97%) | 5.752 (-1.95%) | 4.827 (-1.97%) | 19.15 (-5.77%) | 1.93 (8.43%) | -0.001 (100.37%) | 4.011 (-2.65%) | 32.26 (-7.74%) |

## AUTHOR DECLARATIONS

**Conflict of Interest**

The author has no conflicts to disclose.

**Author Contributions**

**Artem Soshnikov**: Data curation (lead); Investigation (supporting); Methodology (supporting); Software (equal); Supervision (equal); Writing – original draft (lead); Writing – review & editing (equal). **Rebecca K. Lindsey**: Software (equal); Writing – review & editing (supporting). **Ambarish Kulkarni**: Conceptualization (supporting); Methodology (supporting);. **Nir Goldman**: Conceptualization (lead); Data curation (supporting); Investigation (lead); Methodology (lead); Software (equal); Funding acquisition (lead); Writing – review & editing (equal).

## DATA AVAILABILITY

The data that support the findings of this study are available within the article or from the corresponding author upon reasonable request.